\documentclass[a4paper,11pt]{article}
\usepackage{pos}

\title{Neural Network Representation of Generalized Parton Distributions (NNGPD)}

\author*[a]{Simonetta Liuti}

\author[b]{Jitao Xu}

\author[a]{Ho Jang}

\author[a]{Zaki Panjsheeri}

\author[a]{Gia-Wei~Chern}

\author[b]{Yaohang~Li}

\author[a]{Douglas Q. Adams} 

\author[c]{Michael~Engelhardt} 
 
\author[d] {Gary R. Goldstein}

\author[a]{Adil Khawaja} 

\author[e]{Huey-Wen Lin}

\author[a]{Saraswati Pandey} 

\author[a]{Kemal~Tezgin}

\affiliation[a]{Physics Department, University of Virginia, 382 McCormick Rd., Charlottesville, VA 22904, USA}
\affiliation[b]{CS Department, Old Dominion University, Norfolk, VA 22904, USA.}
\affiliation[c]{Department of Physics, New Mexico State University, Las Cruces, NM 88003, USA}
\affiliation[d]{Department of Phyiscs, Tufts University, Medford, MA 02155}
\affiliation{Department of Physics and Astronomy, Michigan State University, East Lansing, MI , 48824 USA.}

\emailAdd{sl4y@virginia.edu}

\abstract{Extracting generalized parton distributions (GPDs) from experimental observables, the convolution structures denoted as Compton form factors (CFFs), requires solving an intrinsic inverse problem. 
We presented results using NNGPD, a flexible regression method based on a physics-informed neural-network framework, for reconstructing GPDs using integral constraints -- including CFFs and Mellin moments -- and symmetry and endpoint constraints. The approach does not require pointwise training information on the underlying GPDs. Closure tests using pseudodata generated from a phenomenological model demonstrate the ability of the framework to reproduce the principal features of the underlying distributions across a range of kinematics. A Bayesian neural-network formulation is also introduced to address uncertainties associated with the reconstruction.
Finally, we discussed a broader question in GPD phenomenology: whether the complete momentum-fraction dependence of GPDs is necessary to extract physical quantities such as partonic angular momentum. This motivates investigating approaches that infer Mellin moments directly from experimentally constrained CFFs, while quantifying the remaining ambiguities.
}

\FullConference{The 33rd International Workshop on Deep Inelastic Scattering and Related Subjects (DIS2026)\\
4 - 8 May 2026\\
Bologna, Italy\\}

\begin{document}
\maketitle

\section{Introduction}
A central goal in quantum chromodynamics (QCD) is understanding the multidimensional structure of hadrons as it emerges from their fundamental quark and gluon degrees of freedom. Generalized parton distributions (GPDs) provide a framework for describing the correlations between partonic longitudinal momentum and transverse spatial distributions, which can be described as hybrid objects between parton distribution functions and elastic form factors. This combination provides an essential ingredient for investigating the substructure of the proton angular momentum in terms of its quark and gluon components.

The extraction of GPDs from experimental measurements has been so far particularly challenging (see Ref.\cite{Liuti:2026jid} for a review of the associated phenomenology). In deeply virtual exclusive scattering processes, such as deeply virtual Compton scattering (DVCS), the measured cross sections and spin asymmetries depend on Compton form factors (CFFs), which are convolutions of GPDs with perturbatively calculable coefficient functions. Consequently, extracting GPDs requires solving a sequence of inverse problems: first determining the CFFs from experimental observables and subsequently reconstructing the underlying GPDs from these quantities.
The difficulties encountered in GPD phenomenology are closely related to those arising in the extraction of ordinary parton distribution functions (PDFs). Traditional PDF analyses, such as those developed by CTEQ \cite{Hou:2019efy} and MSHT \cite{Bailey:2020ooq}, employ flexible analytical parameterizations whose parameters are determined by minimizing a statistical objective function. The Neural-network-based approach, NNPDF \cite{NNPDF:2021njg}, has instead introduced an alternative strategy aimed at reducing the dependence on assumed functional forms.
For GPDs, however, the inverse problem is more stringent. The convolution structure of CFFs permits distinct GPDs to produce identical observables, giving origin to the so-called ``shadow GPDs" \cite{Bertone:2021yyz}. Increasing experimental precision alone does not necessarily eliminate these ambiguities, making the incorporation of additional physical information essential.
At the same time, the additional kinematic dependence of GPDs introduces additional theoretical constraints than in ordinary PDF fitting. These include the polynomiality of Mellin moments, charge conjugation symmetry properties around the parton momentum fraction, $x=0$, and endpoints behavior.
Such constraints provide complementary information that can be exploited in constructing flexible representations of GPDs.

These observations motivated the development of NNGPD \cite{Xu:2026lko}, a physics-informed neural-network framework designed to reconstruct GPDs from experimentally accessible integral quantities and theoretical constraints, without requiring direct knowledge of the distributions as functions of the partonic momentum fraction.
Beyond the reconstruction problem itself, a more fundamental question emerges: ``to what extent is the complete momentum-fraction dependence of GPDs necessary for extracting specific physical properties of the proton?" In particular, quantities such as partonic angular momentum are related to Mellin moments of GPDs. This raises the possibility of developing inference strategies that connect experimental observables directly to the physical quantities of interest, while explicitly accounting for the ambiguities associated with the underlying GPDs.

\section{The GPD inverse problem and the role of physics constraints}
\label{sec:sec2}
\begin{figure}
    \centering
    \includegraphics[width=0.4\linewidth]{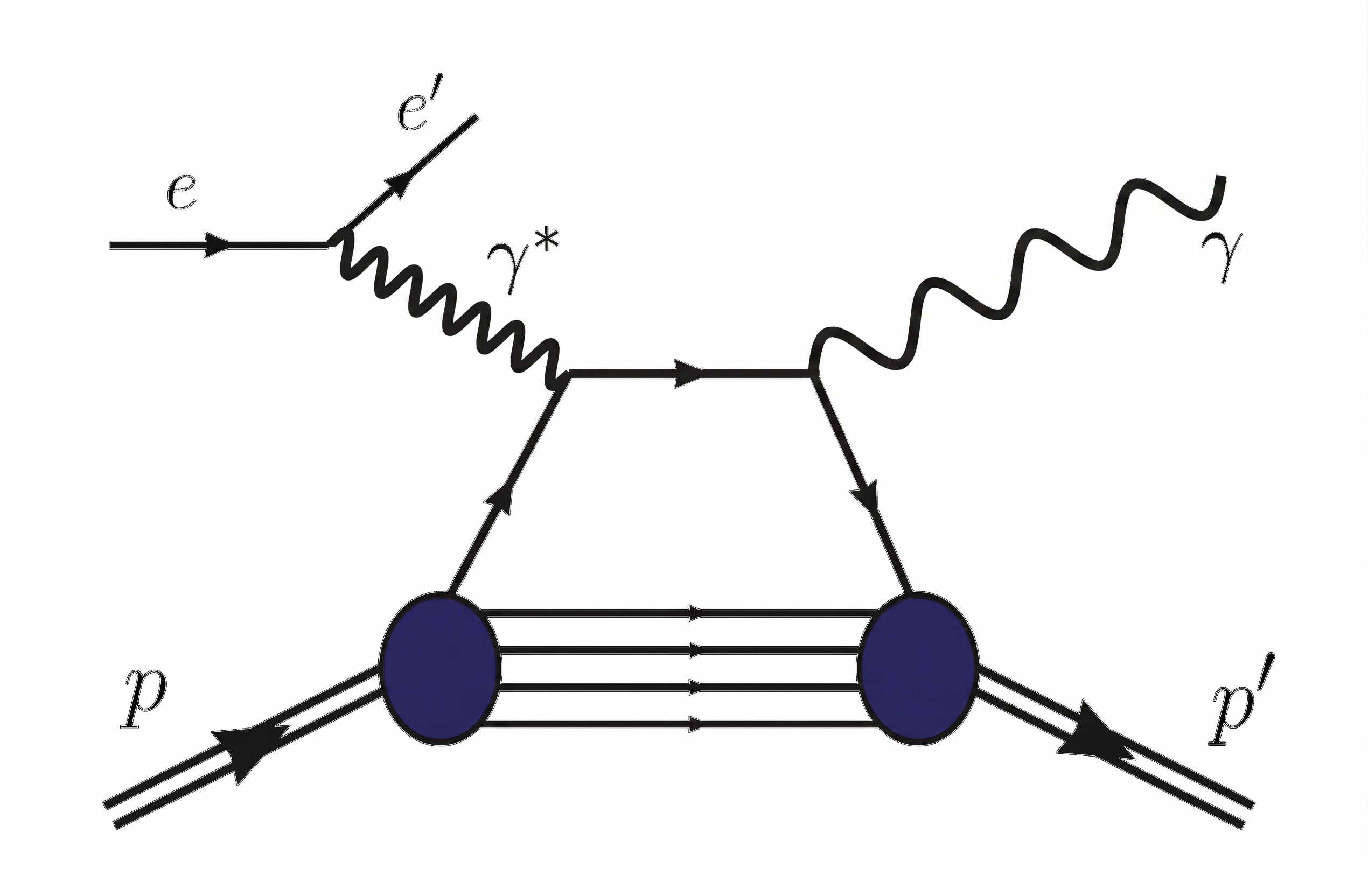}
    \caption{Leading order Feynman diagram for the DVCS process, in the $ep \rightarrow ep'\gamma'$ reaction. The crossed term, $u$-channel diagram, and the background Bethe-Heitler process, where the photon is emitted from the electron, are not shown here.}
    \label{fig:Feynman}
\end{figure}

At leading twist and leading order in perturbative QCD, the observables for a generic deeply virtual exclusive process (Figure \ref{fig:Feynman}), the CFFs can be expressed schematically as,
\begin{equation}
  \mathcal F(\xi,t,Q^2)
=
\int_{-1}^{1}dx\,
C(x,\xi,Q^2)\,
F(x,\xi,t,Q^2),
  \end{equation}
where $F$ denotes a GPD and $C$ is the perturbatively calculable coefficient function. At leading order in perturbative QCD,
the real and imaginary parts of the CFFs can be separated out as,
\begin{eqnarray}
    \mathcal{F}_q(\xi,t)  &=& 
   P.V. \int_{-1}^{1} dx  \left( \frac{1}{x-\xi} + \frac{1}{x+\xi} \right)  F_q(x,\xi,t) 
 {- i \pi \, \big(F_q(\xi,\xi,t) - F_q(-\xi,\xi,t)\big)}\, . 
 \label{eq:CFF}
\end{eqnarray}

The  integral is performed over the partonic momentum fraction $x$, therefore measurements of CFFs at fixed kinematics in the external variables -- $\xi$ (fraction of the longitudinal momentum transfered betweem the initial, $p$ and outgoing, $p'$ protons), $t$ (four-momentum squared transfered between $p$ and $p'$), and $Q^2$, the momentum transfer squared at the electron vertex -- do not directly determine the corresponding $x$-dependence of the GPD. 
Consequently, different GPDs can lead to the same CFF, reflecting the nonuniqueness of the inverse mapping.

The problem is therefore not simply one of insufficient experimental precision, but also of determining which combinations of physical constraints are capable of reducing the space of admissible solutions.
An important source of additional information is provided by Mellin moments,
\begin{equation}
 M_n^q(\xi,t)
=
\int_{-1}^{1}dx\,x^{n-1}H^q(x,\xi,t).
\end{equation}
Lorentz covariance requires these moments to satisfy polynomiality relations in the skewness variable, $\xi$, with coefficients given by generalized form factors. In particular, the $n=1$ moments are connected to the proton's elastic form factors, while higher moments can be constrained by lattice QCD (LQCD) calculations.
An additional constraint arises from charge conjugation which results in the symmetry properties of GPDs under $x \rightarrow -x$. In particular, defining
\begin{equation}
    H_q^{\pm}=(x,\xi,t)\pm H_q(-x,\xi,t).
\end{equation}
Around $x=0$, $H_q^+$, defining the flavor singlet distribution, is anti-symmetric, and $H_q^-$, identified with the valence quarks distribution, is symmetric.
Furthermore, the GPDs are set to zero at the kinematic endpoints, $x=\pm 1$, based on QCD dynamics. These relations provide information about the admissible functional space that do not require accessing information on GPDs at given values of $x$.

The central idea underlying NNGPD is to incorporate these relations directly into the regression problem, allowing the neural network to explore flexible functional representations while restricting the solutions to those compatible with the imposed physical constraints.

\section{The NNGPD framework}
\label{sec:sec3}
NNGPD employs a neural-network representation of GPDs in which the network parameters are determined by minimizing a composite loss function. Rather than training the network on pointwise values of GPDs, the approach uses quantities obtained by integrating the network output, together with symmetry and boundary conditions.
The network is trained using several contributions to the loss function, using the following constraints
\begin{itemize}
    \item Mellin-moments from elastic form-factor constraints and generalized form factors in LQCD, where the appropriate even or odd order of the moment are selected according to the symmetry of the GPD being reconstructed
    \item Symmetry and endpoint constraints: The symmetry properties of $H_q^+$ and $H_q^-$ are incorporated into the training objective, together with endpoint conditions. The constraints also help control the behavior of the reconstruction in regions where the integral observables provide limited sensitivity to local variations of the GPD.
\item CFFs: the loss compares CFFs calculated from the neural-network representation with the input CFF information, $\mathcal H^{\mathrm{input}}(\xi_k,t_k,Q_k^2)|^2.$
Both the real and imaginary parts of the CFF, Eq.\eqref{eq:CFF} contribute to the constraint. 
\end{itemize}
Combining these contributions allows the network to incorporate information from different aspects of the underlying GPD.

\subsection{Bayesian neural networks and uncertainty quantification}
An important question concerns how the residual ambiguity in the reconstruction translates into uncertainties in the extracted distributions. A single optimized neural-network representation does not, by itself, characterize the range of functions compatible with the input information.
To address this issue, NNGPD introduces a Bayesian neural-network (BNN) formulation in which network weights and biases are represented by probability distributions. In the variational approach, the objective includes a Kullback–Leibler (KL) divergence between the approximate distribution of network parameters and a specified prior.
The resulting optimization combines the integral constraints with probabilistic regularization, providing a framework for investigating uncertainties associated with the reconstructed functions.

\section{Closure tests and reconstruction results}
\begin{figure*}
\centering
\includegraphics[width=0.4\linewidth]{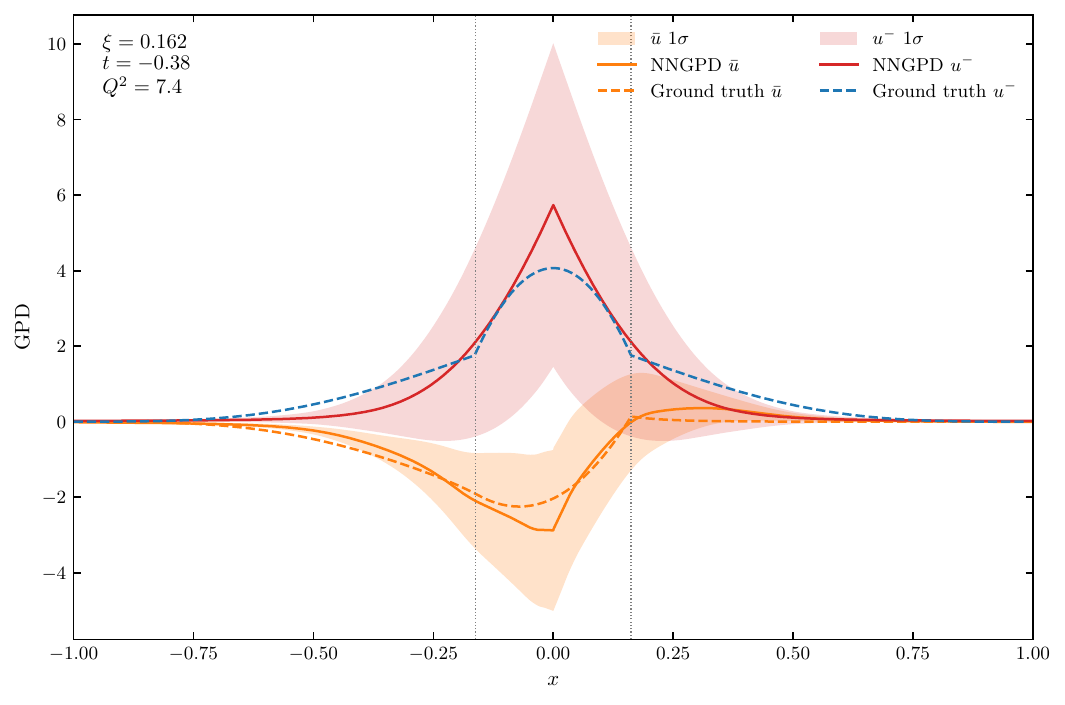}
\caption{(adapted from Ref.\cite{Xu:2026lko}) Example of a comparison of  the {flavor non-singlet, $H_{u_v}(x,\xi,t)$, and anti-quark, $H_{\bar{u}}(x,\xi,t)$}, GPDs, for the kinematics: $\xi= 0.162$, $t=-0.38$ GeV$^2$, $Q^2=7.4$ GeV$^2$, {chosen to cover a kinematic range relevant to experimental measurements}.}
\label{fig:GPD-comparison}
\end{figure*}
The performance of the NNGPD framework is investigated through closure tests using pseudodata generated from a phenomenological spectator-based GPD model.
In these tests, the underlying model provides a known reference distribution against which the neural-network reconstruction can be compared. However, the network is not trained on pointwise values of the reference GPD. Instead, the training information consists of CFFs, Mellin moments, and the additional physical constraints described above.

In Figure \ref{fig:GPD-comparison} we show that NNGPD can reproduce the principal features of the reference distributions over a range of kinematic configurations. The reconstruction is performed separately for the combinations $H_q^+$ and $H_q^-$, $q=u$, which are subsequently combined to obtain the valence quark, $H_{u_v}$ and antiquark, $H_{\bar{u}}$ GPDse.
These results illustrate the potential of incorporating complementary integral constraints into a flexible neural-network representation. In particular, the reconstruction is not determined exclusively by the CFF information, but it benefits from the additional restrictions provided by moments and symmetry relations.

The closure test establishes the feasibility of the constrained-learning approach for the selected benchmark. It does not, however, imply that the inverse problem has a unique solution. The extent to which alternative functions remain compatible with the same input information, and the reliability of uncertainty estimates in regions with limited constraints, remain important questions for further investigation.

\section{Conclusions and outlook}

The NNGPD framework provides a flexible approach to the reconstruction of generalized parton distributions using deep neural networks constrained by physically motivated integral relations. By incorporating Compton form factors, Mellin moments, symmetry properties, and endpoint conditions into a common optimization framework, it becomes possible to reconstruct nontrivial GPD structures without direct pointwise supervision in the momentum fraction.

Closure tests using a phenomenological spectator model demonstrate that the approach can reproduce the main features of the underlying distributions over a broad kinematic domain. These results establish a basis for extending the framework to more realistic combinations of experimental and lattice-QCD information, with particular emphasis on the quantification of reconstruction ambiguities and uncertainties.

More broadly, the development of NNGPD motivates reconsidering the objectives of GPD phenomenology. While the determination of momentum-fraction-dependent distributions remains essential for multidimensional imaging, other physical questions may be addressed more directly through suitably constrained integral quantities.
In this perspective, the success of GPD phenomenology should not be assessed solely by the ability to reconstruct GPDs, but also by the extent to which measurable observables can reliably constrain specific properties of hadron structure, with quantified uncertainties.
This relation raises a question: ``is it necessary to reconstruct the full $x$-dependence of the GPDs in order to determine their moments?"

The conventional approach proceeds by reconstructing or parameterizing the GPDs and subsequently evaluating the required integrals. An alternative strategy would investigate whether the moment information can be inferred directly from experimentally constrained CFFs, supplemented by the relevant theoretical and lattice-QCD constraints.
The crucial issue is whether the available information restricts the desired moment sufficiently, even if substantial ambiguity remains in the underlying $x$-dependent functions.
This distinction is particularly important for the extraction of partonic angular momentum where direct moment inference must account for the theoretical relations and extrapolations connecting the experimentally accessible region to the quantities of interest.
The proposed approach and its future extensions could help establish whether the experimental information is sufficient to constrain a given physical quantity, and identify which additional measurements or theoretical inputs would most effectively reduce its uncertainty.

\vspace{0.5cm}
This work was done under DOE grants DE-SC0016286, DE-SC0024644. 

\end{document}